\documentclass[epj]{svjour}
\usepackage{graphicx}
\usepackage{latexsym}
\usepackage{color}
\usepackage{makeidx}
\begin{document}

\authorrunning{R. Kutner and J. Masoliver}
\titlerunning{Continuous-Time Random Walk still trendy ...}
\title{The Continuous Time Random Walk, still trendy: Fifty-year history, state of art, and outlook}
\author{Ryszard Kutner\inst{1}, 
\and Jaume Masoliver\inst{2}}
\institute{Faculty of Physics, University of Warsaw, Pasteur Str. 5, PL-02093 Warsaw, Poland \and Departament de Fisica Mat\`{e}ria Condensada and Institute of Complex Systems, Universitat de Barcelona, Marti i Franques 1, E-08028 Barcelona, Spain}
\date{Received: date / Revised version: date}

\abstract{In this issue we demonstrate the very inspiring role of the continuous-time random walk (CTRW) formalism and its numerous modifications thanks to their flexibility and various applications as well its promising perspectives in different fields of knowledge. A short review of significant achievements and possibilities is given, however, still far from completeness.} 

\maketitle

\section{Inspiring properties and achievements}\label{section:hypothesis}

By the pioneering work published in year 1965 \cite{MonWeis}, physicists Eliott W. Montroll and George H. Weiss introduced the {\bf concept of continuous-time random walk} (CTRW) as a way to achieve the interevent-time continuous and fluctuating. It is characterized by some distribution associated with a stochastic process, giving an insight into the process activity. This distribution, called pausing- or waiting-time one (WTD), permitted the description of both Debye (exponential) and, what is most significant, non-Debye (slowly-decaying) relaxations as well as normal and anomalous transport and diffusion \cite{KlafZum,ZubDenKlaf} -- thus the model covers essential aspects of the stochastic world -- a real complex world. 

Let us incidentally comment that term `walk' in the name `continuous-time random walk' is commonly used in the generic sense comprising two concepts namely, both the walk (associated with finite displacement velocity of the process) and flight (associated with an instantaneous displacement of the process). Thus it has to be specified in detailed considerations with what kind of process we have to deal.

The {\bf CTRW formalism} was most conveniently developed by physicists Scher and Lax in terms of recursion relations \cite{ScLax0,SchLax,MonWes,WeiRub,MonShle,Weis}. In this context the distinction between discrete and continuous times \cite{Gillis} and also between separable and non-separable WTDs were introduced \cite{HauKehr0}. A thorough analysis of the latter one, called also the nonindependent CTRW, was performed a decade ago \cite{MonMasol} although, in the context of the concentrated lattice gases, it was performed much earlier \cite{KKK,Kut0}. These analysis took into account dependences over many correlated consecutive particle displacements and waiting (or interevent) times.  

The Scher and Lax formulation of the CTRW formalism is particularly convenient to study {\bf as well anomalous transport and diffusion as the non-Debye relaxation} and their anomalous scaling properties (e.g., the nonlinear time growth of the process variance). Examples are the span of the walk, the first-passage times, survival probabilities, the number of distinct sites visited and, of course, mean and mean-square displacement if they exist. It is very interesting that all these things are also using to characterize {\bf complex systems} \cite{ParKanScar}.

In principle, the {\bf CTRW is fundamentally different than the regular random flight or walk models} as probability density of the flight or walk in the long-time (asymptotic) limit scales in a non-Gaussian way, being a serious and inspirational extension of the Gaussian one. Thus, the CTRW became a foundation of anomalous (dispersive, non-Gaussian) transport and diffusion \cite{Weis,ScheMon,PhisScher} opening the modern and trendy segment of statistical physics, as well as condensed and soft matter physics, stimulating their very rapid expansion outside the traditional (Boltzmann-Gibbs) statistical physics (including statistical physics of open systems \cite{EsLin}) \cite{BouGeo,SokKlaf}.

The variety of observed relaxation phenomena in condensed and soft matters are related to transport and/or diffusion most often of atoms, particles, carriers, defects, excitons, and complexes \cite{ZumKlafBlu}. Transport and diffusion are regarded, in fact, as a paradigm of irreversible behaviour of many ordered and disordered systems. A {\bf universal feature of a disordered system} is a temporal complex pattern, which the Debye relaxation no longer obeys. 

The carrier transport in some amorphous insulators (such as the commercially used vitreous $As_2Se_3$) and in some  amorphous  charge-transfer complexes of organic polymers  (as the commercially used trinitrofluorenone mixed with polyvinylocarbazole, TN-PVK) provides canonical examples of (i) continuous-time random flights and walks and (ii) broad- or long-tailed WTDs. The generic description of the {\bf dispersive  transport and diffusion} \cite{MetzKlaft} found in the breakthrough experiments on transient current in an amorphous medium, induced by flash light \cite{ScheMon,Gil,Pfis,Schar,Pfis2} or voltage pulse \cite{CampBradLid}, is given actually by the continuous-time random walk formalism where carrier displacements have instantaneous character that is, carrier performs flights herein instead of walks.

Although originally the CTRW was a kind of renewal theory, Tunaley  was able to modify it by preparing the class of initial (averaging) WTD. Such a modification makes time homogeneous \cite{Tun,Tun1,Tun2,Tun3} and enables to consider CTRW as a semi-Markov process \cite{MeStra,BarLim}. Thus the {\bf application of the key Wiener-Khinchin theorem} (relating autocorrelation function to power spectra) became possible. In other words, there are two categories of initial conditions: the first one, called equilibrated or stationary, where before starting observation of the process, the system is allowed to equilibrate \cite{KlafZumAmst} and the second category, called nonequilibrated or nonstationary, where evolution starts at a given instant without any knowledge of the past \cite{DenZabHang}.  

There is another extremely significant aspect of the above-defined problem of initial conditions, namely the dependence of the formalism on them, which directly leads to {\bf ergodicity breaking} in the Boltzmann sense. We remind that the ergodicity in the Boltzmann sense means that for a sufficiently long time, the time average tends to the ensemble average. One can say, there are two essential interrelated problems of the CTRW: (i) the initial preparation of the system, important for all random walk models and (ii) the weak ergodicity breaking, which is of great interest both from theoretical and empirical points of view.

Diverging of a microscale (or average waiting time diverging) for the non-Debye relaxation and for anomalous transport and diffusion leads to the property of {\bf weak ergodicity breaking} \cite{Bouch,BelBar,MargBark,Heetal,RebBark,RebBark1,LubSokKlaf,Jeoetal,Weietal,ThieSok}. Although then the whole phase space of the system can be still explored, the ergodicity is never obeyed because any measurement time is always shorter or of the order of the time characteristic for considered process. 
This property was introduced in the context of processes showing aging and found experimentally, for instance, in the diffusion of ion channels on the membrane of living cells \cite{WSTK} and in blinking quantum dots \cite{BHMDBD}. 

To say something deeper about {\bf non-ergodic properties} of CTRW processes we should remind that anomalous diffusion loses the universality of Brownian motion that is, the mean-square displacement is no longer sufficient to uniquely identify a stochastic process. Therefore, various stochastic processes give rise to anomalous diffusion exhibiting many different features. For instance, for subdiffusive renewal CTRWs the ergodicity is violated and even weak ergodicity breaking was observed due to the diverging characteristic waiting time \cite{Barkai}.  
Ergodicity breaking effects are essential in understanding the fluctuation generated phenomena, in particular, fluctuation dissipation relations and also linear response \cite{Metzetal}.

Notably, modern empirical single particle tracking techniques and many large scale simulations producing time series of the position of a tracer particle fully confirmed this observation \cite{Metzler}. 
A {\bf non-ergodic generalization of the Boltzmann-Gibbs statistical mechanics} for systems with infinite mean sojourn time was found \cite{BelBar}, which seems to be a great achievement.

It was shown by using extreme value theory (EVT), that indeed rare or {\bf extreme events} actually govern dispersive transport and diffusion \cite{Kut,KozKut}. Thus the CTRW formalism opened promising opportunities to connect a microscopic stochastic dynamics of objects to the macroscopic processes of relaxation, transport and diffusion.

The {\bf biased CTRW formalisms} were also developed, where the bias can affect both on spatial and temporal variables \cite{Tun4,NelHar}. Such an approach constituted a basis for the origin of $1/f$ noise -- it is attributed to the excess noise following from the process fluctuation.  

The appearance of $1/f$ noise can be achieved, for instance, by appropriately modeling trapping states and determining the correspondence waiting-time distribution. The $1/f$ noise can affect several thermally activated processes, e.g., can govern anomalous escape \cite{GoyHang}.

The {\bf biased and non-biased aging continuous-time random walks}, using fractal renewal theory, were prepared \cite{BarYuan}. Thus an essential extension of the canonical (non-stationary) CTRW formalism was developed which contains the aging period prior to relaxing continuous-time random walk. 

The canonical version of the CTRW formalism concerning transitions between different sites and states by using recursion relations is equivalent to the form of {\bf generalized master equation} (GME) as one-to-one transformation between WTD and memory kernel was clearly established 
\cite{KenMonShle,KenMonShle1,KenKno,KenRah,Kenk,LandMonShle,LandShle,LandShle1,Gill,CacWio}. 
Originally, this was used to exhibit the equivalence between models of hopping and multi-trapping in amorphous materials. Notably, kernel without memory corresponds to the ordinary master Markovian equation, while exponentially decaying  memory enables to transcribe GME to a variant of the telegrapher's equation \cite{MasWeis}, when short time mechanism of random walk is activated. 

The {\bf non-Markovian nature of the CTRW} is well seen, in particular, in the presence of disorder in the system considered in the ensemble average. 
The term `non-Markovian' means that the current state of the particle depends on all its history, beyond the recent one.
An equivalence between averaged particle transport in disordered systems and the GME or CTRW formalism was established \cite{KlafSil}. 
Before averaging, the individual disordered systems are Markovian. However, when master equations are averaged over the disorder, by using Zwanzig-Nakajima projection formalism \cite{Zwanz,Haak}, 
it leads to the GME that is, the memory kernel appears as a result of a double average over possible random walk trajectories and over imperfections present in the system. Generally speaking, the vast majority of considered versions of CTRW formalisms (not only those referring to disordered systems) can be treated as renewal semi-Markov processes.

The {\bf renormalization-group (RG) treatment of the canonical CTRW formalism} was performed by using a decimation procedure in one dimension \cite{Mach,Mach1}. 
The resulting fixed-point equation for WTD gave an excellent analytical solution which does not correspond to the Poisson process in a macroscale (although initially, in a microscale, it was so) and, paradoxically, does not depend whether disorder is present in the system or not. This shows that the {\bf CTRW formalism is suited to study systems near criticality} insensitive on a model details.

The {\bf correlated CTRWs} belong to a class of RW where memory is not lost after each step. They were found in various physical applications, e.g., in conformation of polymers \cite{Kuhn,Kuhn1} 
and tracer diffusion in metals \cite{BarHer,Man}. 
The simplest version of these CTRWs assumes correlation over two successive steps, both persistent and antipersistent. Such a model is equivalent with a second-order integro-differential equation \cite{HauKehr,HauKehr1}. 
Notably, in several papers the properties of persistent and antipersistent CTRWs were considered \cite{MaLinWe,Halper}, e.g., in the context of determination of Hurst and H\"{o}lder exponents. Therefore, it seems a natural way to adopt CTRW to study {\bf multifractal types of random walk}. It can be based on the well-known concept to characterize the stochastic phenomena by using a spectrum of fractional moments both temporal and spatial \cite{MetzKlaft,PMKK,KKPM,Castetal,ArtCris,SanLar,deAnetal,Rebenetal,Rebenetal1,SeuStan}. The scaling of the fractional moments of the L\'{e}vy walk has a characteristic fractal, bifractal, and multifractal behaviour. Besides, the common observation was that the diffusion coefficient and conductivity are then frequency dependent quantities which are similar to that observed experimentally, e.g., for ionic conductors. Moreover, the persistent CTRW leads to often used the telegrapher's equation \cite{HauKehr0}. 

A {\bf model of subdiffusion}, where waiting times are mutually dependent (i.e. intrinsically correlated), was also developed by using stochastic dynamic (Langevin) equations. This led to anomalous diffusion under the influence of an external force field \cite{MagMetSzZe}. A very helpful effective theoretical criterion for subdiffusion was found in the context of the reference fractional brownian motion \cite{MWBK}.
It is highly characteristic that subdiffusion has been found in very different systems, beginning with the seminal discovery of charge carrier anomalous transport in amorphous (glossy) semiconducting films mentioned above (in this context the CTRW formalism in the presence of traps was also developed \cite{WeiHav}). Other attractive examples could be tracer dispersion in subsurface aquifers \cite{Scheretal,BerCorDenSche},
hiking on percolating clusters \cite{KleMetKim}
or even  bacteria in biofilms \cite{RogWalWai} 
and of tracers in crowded media, in particular, in living biological cells \cite{GolCox,WebSpaThe,JeoTejBurBarSelBerOdMet,BroIsKepMaiShavBarGar}. 

The CTRW formalism is well suited to study {\bf first-passage time (FPT) problems} related to the trap and escape problems by using, the so-called, survival probability (SP) directly expressed by the WTD \cite{HauKehr0}. Numerous physical applications of these problems exist. For instance, FPT of a randomly accelerated particle, FPT problems in anomalous diffusion, FPT of intermittent random walks, FPT phenomena on finite inhomogeneous networks, FPT of network synchronization, FPT statistics for random walks in bounded domains, FPT behavior of multi-dimensional fractional Brownian motion and application to reaction phenomena.

Exact relations for the path probability densities of a broad class of anomalous diffusion processes were derived by employing the {\bf path integral formulation} \cite{EuFri}. A closed analytical solution for the path probability distribution of a continuous time random walk process was derived. This analytical solution is given in terms of the corresponding waiting time distribution and short time propagator coming from Dyson equation. For instance, by applying this analytical solution, the generalized Feynman–Kac formulae was found.

{\bf A random walk along the backbone over comb} with teeth lengths varying according to power-law distribution, is another interesting example of the CTRW application 
\cite{AvrShl}. 
As a result, the  CTRW predicts anomalous diffusion (subdiffusion) when local WTD (i.e. conditional at given teeth length) is also given by power-law. Anomalous diffusion in comb-like structures serves as a reference model for random walk in more complicated fractal substrates, for instance, such as percolation clusters successfully reproducing the delaying effects of dangling ends and the backbone irregularities in percolation clusters.

Thanks to its versatility, the CTRW found numerous {\bf important applications in many fields} ranging from biology through telecommunication to finance including econometrics and economics, and even to speech recognition. The CTRW found innumerable applications in many other fields, still growing, such as the aging of glasses, a nearly constant dielectric loss response in structurally disordered ionic conductors as well as in modeling of hydrological models and earthquakes. Since the canonical CTRW was first successfully applied by Scher and Lax in 1973 \cite{SchLax} and independently by Moore one year later \cite{Moor} to describe anomalous transient photocurrent in an amorphous glassy  material manifesting the power-law relaxation, this formalism has achieved much more than its original goal. An approachable description of fractional kinetics with characteristic applications to anomalous charge transport and relaxation in solids such as, for example, disordered semiconductor structures, quantum dots and wires, dielectrics (polymers and ceramics), as well as nanosystems, can be found in ref. \cite{UchSib}.

\section{CTRW formalisms vs. fractional evolution equations}

We can say that modern era of the CTRW formalism or its "second youth" was launched in 1999-2000 by series of papers regarding its fractional representations and generalizations as a result of long-term memory kernel or non-Markovian type of underlying processes. Among them the paper \cite{MainRabGorSca} of Mainardi, Raberto, Gorenflo, and Scalas is particularly impressive. They derived (in the frame of the CTRW) an alternative version of the GME and hence the time-fractional Kolmogorov-Feller type equation (see also Sec. \ref{section:FKF2}), considering long-term weakly singular memory that exhibits power-law time decay. In this equation usual partial time derivative was naturally replaced by the Caputo fractional derivative without use of the Riemann-Liouville fractional one \cite{ScaGorMain}. It should be noted that this form of memory kernel implies, e.g., sojourn probability in the form of the Mittag-Leffler function and the corresponding WTD as its ordinary derivative -- it is a promising extension as these quantities also appear in fractional relaxation and fractional oscillation processes \cite{GorMain,Main,Main1}. Notably, the sojourn probability quite well reproduced the dynamics of BUND future prices traded in LIFFE with various delivery dates in 1997 \cite{MainRabGorSca}. That is, the approach mentioned above seems to be very useful for financial analysis.

Moreover, evidences appeared (see paragraphs \ref{section:FDE} -- \ref{section:LevyWalk} below) that various models of fractional stochastic dynamics and kinetics as well as fractional stochastic processes are in macroscopic limit (that is, for the asymptotic long time and for large values of space variable) equivalent to the CTRW formalism. 

\subsection{Fractional diffusion equation}\label{section:FDE}

On three different ways the equation of diffusion can be modified to fractional diffusion equation (FDE) so that to be equivalent to the asymptotic form of the canonical CTRW formalism, i.e. the separable (independent or decoupled) and uncorrelated continuous-time random walk formalism characterized by diverging mean waiting time reflecting the existence of deep traps in the system. This means that average depth of the trap is greater in this case than the Boltzmann constant multiplied by the absolute temperature.

{\bf The first way} leads to FDE describing subdiffusion, where the partial time derivative is replaced by the corresponding Marchaud fractional derivative \cite{Hilfer}. {\bf The second way} leads to FDE where the Laplacian in diffusion equation is completed with the partial time differentiation and with the Riemann-Liouville fractional differintegration \cite{MetzKlaft,MetzKlaft1}. On this way even more general FDE is reached where Laplacian is replaced by the Riesz-Weyl fractional derivative \cite{MetzKlaft}. Thanks to this, the competition between long-lived rests (waiting) of the particle and its long jumps (flights) produces rich phase diagram consisting of four different random walk phases: canonical L\'{e}vy flight (LF), non-Markovian LF, Brownian diffusion, and subdiffusion. Notably, the fractional diffusion-advection equation (FDAE), containing usual advection term, was easily obtained on this way in the laboratory frame of reference. In this case, the Galilei invariance is obeyed between laboratory and moving frame of references, where one-dimensional velocity field is homogeneous. Otherwise, when velocity field is inhomogeneous, the resulting FDAE has the same structure as the fractional Fokker-Planck equation (discussed in the next paragraph).
 
\subsection{Fractional kinetic equation}

{\bf The third hybrid way}, which combines mentioned above approaches, leads to the generic FDE with long-term source containing the initial (static) distribution, where both time and space fractional Riesz/Weyl derivatives were used \cite{ZubDenKlaf,MetzKlaft,SaichZas,Barkai2}. This equation (called also fractional kinetic equation (FKE)) has several interesting regimes covering, besides the case of normal diffusion, also sub- and superdiffusion -- the latter considered also in Sec. \ref{section:LFDE} by L\'{e}vy fractional diffusion equation.

Notably, solutions of the FDEs presented above were obtained in the closed analytical forms but only the second and third ways give the canonical L\'{e}vy distribution as a solution (for proper combination of scaling exponents) \cite{ZubDenKlaf}. 

\subsection{L\'{e}vy fractional diffusion equation}\label{section:LFDE}

The L\'{e}vy flights (LFs) belong to the class of Markov process with broad flight length distribution possessing asymptotically the inverse power-law behaviour such that its variance diverges. This scale-free `area-unfilling' or fractal nature of this distribution gives, in force-free case, the L\'{e}vy fractional diffusion equation (LFDE) \cite{SaichZas,Comp,Fog,Honk}. In this diffusion equation the fractional Riesz/Weyl derivative replaced the Laplacian and fractional diffusion coefficient appeared instead of the usual diffusion coefficient. The time-dependent solution of the LFDE has a closed form which can be represented by the Fox $H$-functions \cite{MetzKlaft,Jespetal}. Obviously, this solution gives well-known L\'{e}vy stable law.   

Due to the Markovian character of the LFDE, the constant drift velocity can be easily incorporated into the LFDE in the form of usual drift or advection term \cite{MetzKlaft,MetzKlaftSok}. This is possible if mean waiting time is finite. This equation can be called the L\'{e}vy fractional diffusion--advection equation (LFDAE). The closed solution of this extended equation is given, in fact, by that for the free case with space variable shifted by displacement caused by the drift. 

\subsection{Distributed-Order Fractional Diffusion Equation}

The anomalous non-scaling behavior, corresponding either to a non-power-law (e.g., the logarithmic) growth of a distribution width or to a crossover between different power laws, are observed quite often. A commonly known examples of such a behavior serve the Sinai-like superslow subdiffusion or superdiffusive truncated L\'{e}vy flights -- all of them can be described by diffusion equation with distributed-order (fractional) derivatives \cite{ChSoKl}. We are dealing with the distributed-order derivative when average over orders of its corresponding fractional components is made. Two generic cases should be clearly distinguished herein: (i) distributed-order time fractional diffusion equation and (ii) distributed-order space fractional diffusion equation. For both cases (implemented through their different forms) the weights play a central role and relations to the CTRW is well established.

\subsection{Fractional Telegrapher's equation}

In the context of diffusion theory, the telegrapher's equation (TE) is seen as a relativistic generalization of the diffusion equation, since the latter is not compatible with relativity \cite{schlesinger_95,keller,abdel,dunkel,maso_weiss96}. It also takes into account ballistic motion and tends to be more accurate in modeling transport near boundaries than the diffusion equation \cite{weiss_02}. Within the surge of anomalous diffusion during the last two decades, there have been some (few) attempts to generalize TE to include fractional motion. Thus, in the mathematics literature, there have been some works  analyzing mathematical and other formal properties of a fractional version of the TE. However, the fractional equation is set in an {\it ad hoc} fashion, just by replacing ordinary derivatives by fractional derivatives that can be of various types \cite{or_03,or_04,or_14}. Efforts meant to derive the fractional telegrapher's equation (FTE) based on physical grounds are very scarce \cite{compte_97,compte_99,metzler_98}. Although, very recently, it has been presented a rather thorough derivation of the one-dimensional FTE \cite{masoliver_16} based on the (fractional) persistent random walk --a variant of the CTRW which allows for the presence of internal states and that incorporates a form of momentum within the framework of diffusion theory \cite{weiss_02,katja_89}.

\subsection{Fractional Kolmogorov-Feller equation}\label{section:FKF2}

The well-known Kolmogorov-Feller equation is somehow the master equation used in different physical applications (see the first paragraph of this section). However, further generalization is required which enables to describe the non-Markovian kinetics \cite{SaichZas}. This was achieved by (i) replacing partial time derivative in the (usual) Kolmogorov-Feller equation by Riesz/Weyl one and (ii) extending this equation by the source term localized at the origin and slowly decaying in time.

\subsection{Fractional master equation}

There is also a fractional generalization of the conventional master equation for the subdiffusive random walk (initially located at origin), where partial time derivative is replaced by the fractional one \cite{HilfKutPekSznaj}. This generalization  
becomes the canonical CTRW formalism in the macroscopic limit by assuming the WTD which asymptotically exhibits a power-law tail. 

\subsection{Fractional Fokker-Planck equation}\label{section:FFPE}

A framework for treatment of anomalous diffusion problems under the influence of an external force field is herein presented. This is a response to the demand arising from the fact that many transport and diffusion problems in science and technology take place under the influence of an external force field \cite{MetzKlaft}, sometimes near the thermal equilibrium \cite{BarMetzKlaf}. 

The effort was made by many authors to find the most general circumstances defining the above mentioned framework. A particularly inspiring, in some sense more general than the FDE reference to the CTRW, supplies a one-dimensional nonhomogeneous fractional kinetic equation or fractional Fokker-Planck-Kolmogorov equation (FFPKE) describing a symmetrized in space wandering with a pointwise algebraically relaxing source \cite{SaichZas}, 
where both time and space derivatives are fractional. 

For description of anomalous transport in the presence of an external field, the fractional Fokker-Planck equation was found, where the Fokker-Planck operator was completed with the partial time differentiation joint with the Riemann-Liouville fractional differintegration \cite{MetzKlaft,MetzKlaft1}. Its stationary solution is given (as for the usual FPE) by the Gibbs-Boltzmann distribution. For the force-free limit the FFPE reduces to the FDE mentioned in Sec. \ref{section:FDE}.

Taking also a non-local jump statistics into account, i.e. assuming a jump length distribution with infinite variance, one recovers generalized FFPE where Laplacian in the Fokker-Planck operator was replaced by the Riesz/Weyl fractional derivative. Thus it describes the competition between subdiffusion and L\'{e}vy flights. Obviously, if order of the Riesz/Weyl fractional derivative equals 2, the generalized FFPE reduces to the subdiffusive FFPE mentioned above. Notably, the FFPE enables to study subdiffusion under influence of time-dependent alternating force fields or driving \cite{HPGH,HPGH1}.

Assuming the finite mean waiting time, the Markovian FFPE can be derived for L\'{e}vy flights, which is the usual FPE where only Laplacian is replaced by the Riesz/Weyl operator \cite{MetzKlaft}. Such a FFPE describes systems far off thermal Boltzmann equilibrium. This is well seen for the behaviour of a particle underlying the harmonic potential, where stationary state is defined by the L\'{e}vy stable law.

Moreover, the FFPE which contains a variable diffusion coefficient, is discussed and effectively solved \cite{Srok}. It corresponds to the L\'{e}vy flights in a nonhomogeneous medium. It is interesting that for the case with the linear drift, the solution becomes stationary in the long-time limit representing the L\'{e}vy process with a simple scaling.

\subsection{Generalized Langevin subdiffusion dynamics}

There are two characteristic, essentially different mechanisms underlying the subdiffusive dynamics that is, subdiffusion in presence of a tilted washboard (nonlinear) potential energy profile \cite{GoyHan}. The first approach is based on the fractional Fokker-Planck equation (derived within the continuous-time random walk), while the second approach is associated with the fractional Brownian motion in the form of a generalized Langevin equation (GLE) \cite{ErLut} that is, containing memory-friction slowly relaxing term. Indeed, for such a potential the difference between both approaches become particularly distinct. For instance, it was found that the second approach is a more ergodic (asymptotically ergodic) than the first one as the latter is based on the concept of fractal stochastic time with divergent mean period while with finite mean residence time in a finite spatial domain. Moreover, the anomalous transport coefficient became universal within the GLE, obeying the generalized Einstein relation (in the absence of periodic potential). Remarkably, the GLE subdiffusion is based on the long-range velocity--displacement correlation and not on diverging mean residence time in a potential well -- the latter case is closely tightened to a weak ergodicity breaking. Hence, the contrast to the CTRW subdiffusion (with independent increments) clearly arises showing that both approaches belong asymptotically to different universality classes. Concerning applications, the GLE subdiffusion dynamics seems to be appropriate, e.g., for regime slightly above the glass transition or for crowded viscoelastic environments (like cytosols in biological cells).

\subsection{Fractional Klein-Kramers equation: subdiffusion}\label{section:subdiffus}

The dynamics in phase space spanned by velocity and position, governed by the multiple trapping process (both subdiffusive and Markov limits) \cite{MetzKlaft,MetzKlaft1}, is defined by the fractional Klein-Kramers equation (FKKE). The Klein-Kramers operator is in this equation completed with the partial time differentiation and with the Riemann-Liouville fractional differintegration. Note that the Stokes operator \cite{Chandra} is replaced in the FKKE by the corresponding fractional one which shows the non-local drift response due to the trapping \cite{MetzKlaft1,MetzKlaft0,MetzKlaft2,Metz1}. 

One can consider the under- (velocity equilibration) and overdamped (large friction) limits. The former limit corresponds to the fractional version of the Rayleigh equation in the force-free limit \cite{MetzKlaft0,MetzKlaft2,vanKamp}. It is a subdiffusive generalization of the Ornstein-Uhlenbeck process. In the overdamp case, the FKKE corresponds in position space, to the FFPE \cite{MetzKlaft0,MetzKlaft2,Metzetal1,Metzetal2}. In this case, the initial condition is persistent due to the slow decay of the sticking probability. Besides, generalized Einstein-Stokes relation and linear response in the presence of a constant field are obeyed.

\subsection{L\'{e}vy fractional Klein-Kramers equation: superdiffusion}

There are several ways to obtain L\'{e}vy fractional Klein-Kramers equation concerning superdiffusion in phase space spanned by velocity and position of a single flyer \cite{Metz1,Lutz,Pes,Kus,Lutz1}. This equation is a generalized Klein-Kramers equation where the second-order partial derivative attached to the flyer velocity is replaced by the fractional Riesz/Weyl derivative. The L\'{e}vy FKKE allows the divergence of the flyer's kinetic energy \cite{SesWes} as a result of the linear (hydrodynamic) too small friction inherent in this equation. Indeed this friction is a source of unphysical L\'{e}vy flights. These flights can be regularized by permission of the nonlinear (aerodynamic) friction, e.g., ballistically depending on the flyer velocity \cite{MetzKlaft1}. It should be added that the velocity average of the L\'{e}vy FKKE reduces it to the L\'{e}vy FFPE mentioned in Sec. \ref{section:FFPE}.

There are also other ways to regularize the L\'{e}vy flights. For instance, particularly natural is the way which uses a non-separable CTRW, where hierarchical spatio-temporal coupling was exploited \cite{Kutner}. By term `hierarchical spatio-temporal coupling' it is understood a coupling between single-step displacement and preceding its waiting time separately on each level of the hierarchy of waiting-time distributions, extended over infinite many scales or levels. The finite mean-square displacement (MSD) was achieved then for arbitrary time thanks to the competition between flights and waitings -- this competition produced a rich phase diagram (see Fig. 3 in ref. \cite{Kutner} for details), where (combined) diffusion exponent characterizes many diffusion phases defined by partial (i.e. spacial and temporal) scaling (or shape) exponents. Besides the normal diffusion phase, there are subdiffusion, enhanced diffusion, and ballistic diffusion phases. The latter phase defines border, separating these phases from pure L\'{e}vy one (characterized by diverging MSD). It is worth to notice that even if extremely long flights are very likely (e.g., were drawn from L\'{e}vy stable law), the long waiting can compensate them after many steps resulting, for instance, even in subdiffusive CTRW. 

By the way, there are two characteristic models using spatio-temporal coupling, called the CTRW with "jump first" \cite{Zabur1} and CTRW with "wait first" \cite{SKW,Bar1}, which clearly show the influence of the first state on the particle dynamics. The former model assumes the particle jump as a first state while the latter one assumes the waiting instead of jump. This is the only difference between these two models which, however, leads to a distinct difference of the last step at given time t. However, this difference is crucial. Although propagators of both models have the same scaling properties their shapes are model specific. It is worthy to mention that trajectories of the particle produced within these models resemble that of the standard L\'{e}vy walk model.

Moreover, the spatio-temporal coupling was similarly used in regularization of more complex processes, like L\'{e}vy walks   \cite{Kutner1} which have richer phase diagram (cf. Fig. 2 in this ref.). It should be added that both L\'{e}vy flights and L\'{e}vy walks are observed in a real world \cite{Katetal}.

\subsection{Fractional Feynman-Kac equation}

The fractional Feynman-Kac equation (FFKE) is a convenient tool to study several characteristic functionals of the subdiffusive CTRW processes both in the absence and presence of a binding force field (e.g., the harmonic field) \cite{CarmBar}. In the latter case the route to weak ergodicity breaking was shown. The FFKE can be obtained from the usual (integer) Feynman-Kac equation by insertion of a substantial fractional derivative operator instead of the Laplacian and generalized diffusion coefficient instead of the usual one.

\subsection{L\'{e}vy walks}\label{section:LevyWalk}

When particle performs CTRW of the flight type, then during its evolution it makes instantaneous jumps alternated with waiting events or rests. The CTRW formalism enables to combine both particle states, offering an abundant diffusion phase diagram or several scaling regimes. Moreover, the CTRW formalism can be extended assuming walks with finite fixed velocity instead of the instantaneous jumps. Such a kind of model is called the L\'{e}vy walk interrupted by rests \cite{KlaftZum1,ZabChuk,KlafSok}. Although the presence of finite particle velocity there significantly increases the flexibility of this kind of models, it simultaneously makes more difficult a finding of their analytical solutions, if it exists. Obviously, the standard versions of L\'{e}vy walk model, i.e. without rests, were also intensively developed assuming fixed particle velocity \cite{KlafBluShl,ZumKlaf} or varying, e.g., according to the self-similar hierarchical structure \cite{KutSwit}. 

There are several generalizations of the L\'{e}vy walk model assuming that particle velocity can vary randomly \cite{BarKlaf,ZabSchSta} or by some other rules \cite{ZabSchSta}. Among them, L\'{e}vy walk model with random velocity and particularly the one with weakly fluctuating velocity caused by the active environment \cite{ZabDenHang,DenZabHang,ZabDenHang1} are very instructive and useful. In the frame of the former model, each displacement has its own velocity drawn from given velocity distribution. Because of the additional complexity added through velocity distribution, there were found only few analytically solvable examples. Among them the Lorentzian or Cauchy velocity distribution offers the prominent one. This velocity distribution appeared, for instance, in: (i) physical problems of two-dimensional turbulence \cite{TonGol,MML,Chuk}, (ii) as a model distribution of kinetic theory \cite{BMM,TBE}, (iii) as a particular case of generalized kappa distributions of plasma physics applications \cite{HMD,MTS}, and (iv) in some statistics \cite{Tsal,Tsal1}. Moreover, it was also found for the distribution of velocities of starving amoeba cells \cite{Taketal}. In the case of the latter model, particle velocity fluctuates around a fixed averaged value. As a result the fluctuations accumulate with time and the final position of the particle, passing through the active medium, will differ from that produced by the standard L\'{e}vy walk.

Concluding this extremely important section, we can say that finite velocity of walking particles constitutes random walk models more general than the random flight (jump) ones and bring them closer to the physical principles making them more suitable for description of real-life phenomena.

\subsubsection{Useful tools and selected applications}

One of the central question is to understand how the L\'{e}vy walk emerges in diverse phenomena. Certain progress in this respect was achieved thanks to development of necessary tools relevant for the L\'{e}vy walks. Among them the propagator of the single particle process and the space-time velocity autocorrelation function for such a process together with its two-point generalization are particularly useful. Moreover, further extension of the space-time velocity correlation function to the broader class of initial conditions is also possible. It should be noted that all these functions perfectly characterize diffusion phases both normal and anomalous. For many versions of the CTRW formalisms they can be calculated in the analytical closed forms at least asymptotically.

There are also exploited, in the context of L\'{e}vy walks, the complementary quantities such as non-normalizable densities (caused by some singularities) making possible to calculate moment diverging within the standard L\'{e}vy distribution \cite{Rebenetal,Rebenetal1}.

Anyway, the extension of L\'{e}vy walk models to higher dimensions is at the beginning stage of development as even extension to two dimensions encounters difficulties.

\subsubsection{Chaotic advection, chaotic Hamiltonian dynamics, and L\'{e}vy walks with rest}

Chaotic Hamiltonian systems, with their unique properties like cantori and stickiness phenomena, are extremely useful for two-dimensional chaotic advection \cite{Arefetal}. The chaotic advection is the field at the intersection of fluid mechanics and nonlinear dynamics, which encompasses a range of multiscale applications ranging from micrometers to hundreds of kilometers, including systems as diverse as mixing and thermal processing of viscous fluids, micro-fluidics, biological flows, and large-scale dispersion of pollutants in oceanographic and atmospheric flows.

Two-dimensional velocity field of a passive scalar (tracer particle) within the incompressible flow can be expressed by the scalar stream function. That is, coupled equations describing the motion of the tracer particle (tracer dynamics) looks like the canonical Hamilton equations in classical mechanics with the stream function playing role of the Hamiltonian, while both space coordinates of the passive scalar looks like the conjugate coordinates. If the stream function (Hamiltonian) is time-dependent, the passive scalar (phase space point) trajectories can be chaotic. Indeed, this is the chaotic advection which can occur even if the flow is laminar \cite{SoWeSwi,WeSoUrHa}. This was observed experimentally in a rapidly rotating annular tank where flow consists of an azimuthal chain of stable vortices sandwiched between two azimuthal inner and outer jets. If the flow has, e.g., periodic time dependence in the annulus reference frame (and can be even time independent in the reference co-rotating with the vortex chain) the tracer typically follow chaotic trajectories alternately sticking near the vortices and occasionally walking ballistically in jet regions for long distances. The competition between sticking and walks can give rise to anomalous diffusion. Probability distribution functions are measured both for sticking and walk times resulting in power-laws. It was also discovered that the probability distribution function of azimuthal walk lengths is the L\'{e}vy one and the long-term process as a whole can be considered as the L\'{e}vy walk with rests. A very inspiring relation between L\'{e}vy walks with rest of a tracer particle and strange kinetics was discovered in 1993 year \cite{ShleZasKlaf} -- the above presented a real experiment of Solomon-Weeks-Swiney \cite{SoWeSwi,WeSoUrHa} just confirmed it. By term 'strange kinetics' it is understood a kinetic description of a dynamical system exhibiting chaotic behaviour, where nonlinearity in the Hamiltonian can induce fractal motions with exotic statistical properties.

\subsubsection{Applications in optics}

We present herein chosen exciting applications of L\'{e}vy processes (flights or walks) to model {\bf scattering of light by media}. It is, in fact, a multiple scattering by medium inhomogeneities ruled by very different scattering mechanisms depending on the characteristic size and structure of inhomogeneities \cite{ZubDenKlaf} (and refs. therein). For instance, the path of a photon inside a fractal media can be represented by a random walk trajectory consisting of undisturbed segments connecting subsequent scatterers -- the statistics of segments' lengths is a power-law \cite{DaMa}. As photons move with finite velocity in any medium, then the L\'{e}vy walk model is more appropriate to describe the photon dynamics than the L\'{e}vy flight although, in some experiments, e.g., type of the light transmission in the L\'{e}vy glass, the latter process gives already a sufficient description. 

The {\bf blinking quantum dots} (QDs)  is another modern example which can be mapped to the L\'{e}vy walk in the ballistic regime \cite{Nirmaletal,Margolinetal}. It is speculated that the L\'{e}vy walk model with random velocities could be useful for the interpretation of experiments with a whole distribution of intensities. Nevertheless, the microscopic mechanism responsible for the appearance of the power-law distributed blinking times in quantum dots remains unknown.

L\'{e}vy flights and walks proved to be very useful, still developing, theoretical tools in the field of {\bf cold atom optics}. The former appeared as an anomalous diffusion related to subrecoil laser cooling process \cite{Aspectetal,Bardetal,Reichetal}. The latter appeared in the context of Sisyphus cooling of atoms loaded into an optical bipotential created by two counterpropagating linearly polarized laser beams \cite{DalCoh-Tan,DechShaKesBar}.

\subsubsection{Search and foraging strategies}

Searching and foraging is now located in the main stream of ecology arousing an increasing interest of researchers. This is a young area of research inspired by pioneering papers on the L\'{e}vy flights of albatrosses \cite{Viswanetal} and on the optimality of the L\'{e}vy walk search \cite{Viswanetal1}. This area contains extremely complex problems (e.g., the animal search) which depend on many irreducible unpredictable significant factors. Although the validity of the concept of the     L\'{e}vy processes is still controversial in the ecological context, several comprehensive monographs \cite{Benietal,Viswanetal2,MCB} support its spreading even beyond animals to humans and robotics.

Recently, L\'{e}vy walks became useful for communities researching the multiscale search and foraging strategies, and motility of living organisms from very primitive to extremely complex ones. It is a challenge to quantify their behaviour as it spans many scales (ranging from swimming bacteria to albatrosses which can soar even for hundreds of kilometers). Living organisms involve complicated interactions with environment containing their habitats. Nevertheless, L\'{e}vy walks are involved in the question of effectiveness of motility in the context of search and foraging strategies.
  
\section{Financial applications of CTRW}

\begin{figure}
\begin{center}
\bigskip
\includegraphics[width=85mm,angle=0,clip]{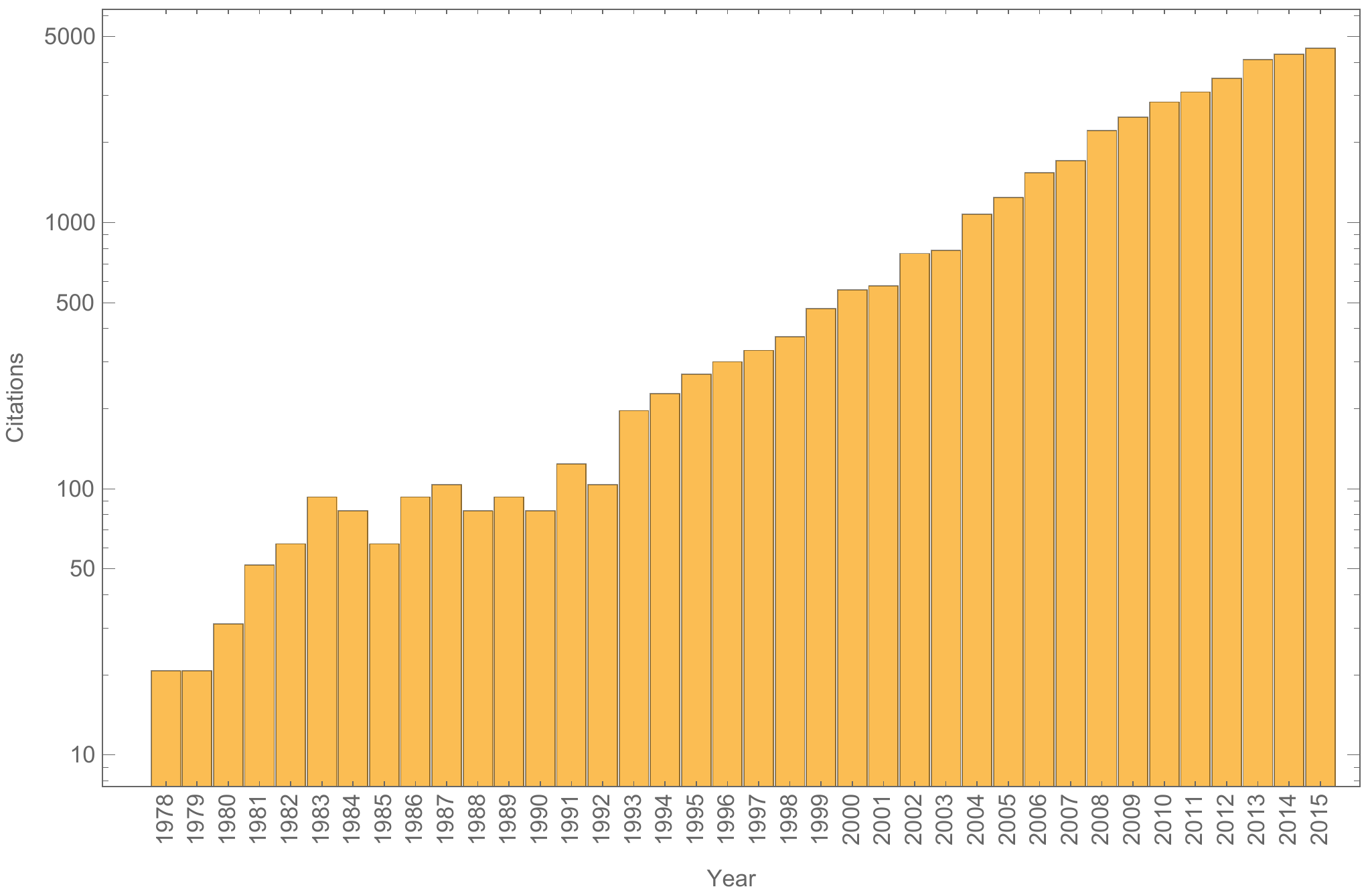}
\caption{A report taken from Web of Science from the years 1978 (i.e., the year when the first review \cite{PhisScher} of CTRW appeared) to 2015 (i.e., fifty years after publication of the Montroll and Weiss CTRW). Results were shown in the semi-logarithmic plot (sum of Times Cited without self-citations: 32333. H-index: 91). Notably, the exponential growth of citations is well seen.}
\label{figure:citations}
\end{center}
\end{figure}

Multidirectional applications of different versions of the CTRW formalism is a research trend which developed quite quickly in the last decade resulting in an exponential growth of citations (see Fig. \ref{figure:citations} for details). A prominent example of going far beyond the traditionally understood physics is an application of the CTRW formalisms by econophysicists that is, in economics and finance -- their achievements have been presented in the review by Enrico Scalas in 2006 year \cite{Scal} and also in refs. \cite{SGM,MMW,MMPW,KYKLS}. Since then, a number of hopeful proposals appeared for the use of different variants of the CTRW to describe multiscale statistical properties of assets (e.g., their dependencies, memory, and correlations) quoted on financial markets \cite{PMKK,TGRK} (and refs. therein). Very recently, some universal properties or superscaling of superstatistics (i.e., scaling of scaling exponents) of returns for diverse financial markets were described by such a version of the renewal CTRW formalism, which contains thresholds separating explicit state of the walker activity from its hidden state \cite{DGKJS}. It is not very surprising that this formalism well reproduced threshold empirical seismic activity data concerning various earthquakes.

\section{Concluding remarks}

The continuous-time random walk models define, in fact, two-state formalisms where the active state is defined by the walker displacements while the passive state is nothing else than waiting. These two essentially different states are present alternately during the journey of the walker. Obviously, one deals with extreme versions of the CTRW formalisms when the passive state of the walker vanishes at all or its displacements in the active state are instantaneous. The canonical CTRW formalism expanded for surprisingly many formulations and versions, where the canonical (original) version is only a very special case. 

This year begins the next half-century of the continuous time random walk's mature life. The enclosed statistics presents citations in each year together with a short report (taken from Web of Science). This confirms the incessant, great interest in the possibilities of CTRW methodology.

The present issue contains a collection of carefully selected papers on various deep and rich aspects of the CTRW and its promising inspirations prepared by scientists prominent in the field. 

\end{document}